# Impact of State and State Sponsored Actors on the Cyber Environment and the Future of Critical Infrastructure

Henry Durojaye
*Cybersecurity & Human Factors*
*Bournemouth University,*
Bournemouth, United Kingdom
S5526113@bournemouth.ac.uk

Oluwabukola Raji
*Cybersecurity & Human Factors*
*Bournemouth University,*
Bournemouth, United Kingdom
S5526171@bournemouth.ac.uk

*Abstract*—**The purpose of this research paper is to critically explore the impact of state and state-sponsored actors on the cyber environment and the future of critical infrastructure, the majority of these attacks on the cyber environment have focused more on the vulnerability of critical infrastructures, this can be evidenced in the cyber-attack by Russia on December 23rd, 2015 that caused power outages experienced by the Ukrainian power companies which affected many customers in Ukraine [8]. Considering the enormous resources available to state and state-sponsored actors it has become difficult to detect cyber-attacks, even when the attack is discovered, proving that it was carried out by a particular state is not easy as such it is now being commonly exploited by malicious states. The paper examines the effect of the actions of the state and state-sponsored attacks on the cyber environment and critical infrastructures, these adverse effects include; greatly diminished defense capacity of the attacked states, destabilise the micro-economy, disinformation that can effectively sway public opinion in a state [1]. Consequently, being aware of the number of resources available at the disposal of the actors and the enormous negative impacts on the cyber environment and critical infrastructures, the government, agencies, and other professionals will be prepared to protect and prioritise network and security systems as a national issue thus encouraging public-private collaboration.**

*Keywords — Critical infrastructure, cyber warfare, state actors, state-sponsored actors, cyberattacks.*

## I. INTRODUCTION

As a result of the continuing digital transformation trends, there has been a rising emergence of a data-driven society where most actions are being informed by data [1]. When this is contextualised with the emergence of the digital era that has resulted in the increased focus on improved operational efficiency, governments have resorted to the development of critical infrastructure to direct and improve service delivery. As such, critical infrastructure refers to the assets developed and maintained by governments that are critical to the functioning of an economy and society. While this has contributed to improved and efficient delivery of government services, critical infrastructure is gradually becoming a target for threat actors. Given the growing importance of critical infrastructure to societies and economies around the world, it is imp to improve critical infrastructure protection [2]. This is especially apparent when considered from the perspective that threat actors are employing more advanced tactics, techniques, and procedures to infiltrate networks and systems. In addition to this, attacks are increasingly becoming more targeted which is contributing to the increase in their severity and chances of success [3]. To this, the entry of state and state-sponsored threat attackers foreseeably increases the susceptibility of governments to disruptions especially from the context of the rising rates of cyber warfare between countries and gangs coupled with the relative low likelihood of being incarcerated.

### A. Problem statement

The attacks against the US government networks and US private companies allegedly by hackers backed by Russia and China have resulted in the increased attention on state-sponsored attacks. In mid-march 2021, the Center for Strategic & International Studies estimated there were 20 such attacks. Among the attacks include the attack by the Chinese government on Microsoft Exchange Server users and the SolarWinds hack by Russia. In the aforementioned data driven society, state and state-sponsored threat actors are motivated by furthering a state's agenda which includes seeking control or disrupting infrastructure and other core systems and information held by another country's organisations, agencies, and institutions [4]. For the different states, state and state-sponsored groups are preferred to conduct cyber warfare since these groups tend to have better operational security and they focus on acting covertly to attain their objectives. Therefore, as governments increasingly become reliant on critical infrastructure in the delivery of essential services including improving governance outcomes, state and state-sponsored actors pose a foreseeably key challenge. This way, evaluating their impact can help governments better safeguard their systems and networks.

### B. Significance of the research

The Covid-19 pandemic has contributed to a rise in geopolitical tensions, which have been furthered by the Ukraine/Russia conflict [5]. In a joint Cybersecurity Advisory (CSA), the governments of Australia, Canada, the US, New Zealand, and the United Kingdom have warned organisations that the invasion of Ukraine by Russia can contribute to the exposure to increased malicious cyber activity. This is being contributed to from two main fronts. The first one is that Russia is exploring cyber warfare as a key attack vector to which they have deployed attacks such as distributed denial of service attacks and deploying malware on the Ukrainian government's critical infrastructure [3]. The second one is that some cybercrime groups have publicly pledged to support the Russian government. As such, they have threatened to carry out cyberattacks in retaliation for the perceived offenses against the Russian government and its people. In addition to just targeting the Ukrainian government, these groups have also threatened allies to the Ukrainian government [3]. Therefore, in line with the suggestion by different cybersecurity authorities in different countries, it is important for different governments to prepare and mitigate against potential cyber treats by reinforcing their cyber defenses and employing the necessary due diligence.

*C. Research aim and objectives*

Owing to the increased preeminence of data driven societies and the increased attractiveness of critical infrastructure to threat actors, the aim of this research is to evaluate the foreseeable level of victimisation of critical infrastructure to state and state-sponsored threat actors.

*1) Research objectives*

To attain the aforementioned research aim, this study will be directed by the following objectives:

- To establish the commonly used modes of attack by state and state-sponsored threat actors.
- To outline the factors that contribute to the increased attractiveness of critical infrastructure to state and state-sponsored threat actors.
- To determine the extent to which critical infrastructure is being targeted as a form of cyber warfare.
- To outline the core considerations and priorities that governments and government agencies should apply in security systems and networks.

## II. REVIEW OF LITERATURE

*A. Politicisation of the cyber environment*

For most parts of the world, societies rely on uninterrupted access to digital technologies to access diverse critical essential services. It has emerged that attacks are gradually becoming more targeted and impactful and for most cases more political and strategic. This is outlined to be contributed to by the increased interlinking of technology, politics, research, and science [6]. In line with this, threat actors are increasingly using or misusing digital technologies in political and socioeconomic context, and across formal and informal settings between a state and its bureaucracies to define roles, acceptable code of conduct, boundaries, and responsibilities. As such, this raises the questions regarding the political aspects and the politics of engaging with the question of securing cybersecurity for critical infrastructure. This goes to show that as countries around the world seek to adopt more advanced technologies to direct service delivery, the interconnectedness between complex socio-technical systems is projected to only grow exponentially [6]. However, it is important to understand that digital technologies have politics ingrained in them while technological possibilities and developments require the development of new governance mechanisms which are also shaped by politics. Therefore, it is evident that as communities and economies around the world continue to adopt digital technologies, there is an emergent hierarchy between different entities at the society level and globally.

As such, it emerges that globally, the internet has become more unsafe gradually and the threat has moved from individuals to the national levels. When this is contextualised with the entry of state actors, there has emerged a contested cyber space where aspects such as economic espionage, intelligence, psychological operations, and information operations have radically transformed key considerations for internet security [7]. Therefore, while the years following the development of the internet were characterised with being satisfactorily secure because of the limited abilities of attackers, the entry of state and state-sponsored actors has resulted in the relative increase in expertise and knowledge among cyber criminals. Also, the increased targeting of cybersecurity campaigns on national economies is contributed to the increased severity of attacks as attackers seek to exploit weaknesses within a national infrastructure, industry linkages, and psychological operations [6]. What is more, cybersecurity attacks have an inherent weakness of attribution which makes them effective for diverse cybersecurity attacks by state and state-sponsored attacks such as proxy wars or utilising criminal networks or political groups to carry out attacks. As such, this is contributing to the increased contribution of cyberspace on national security.

Finally, the role of a state in cybersecurity needs a diverse approach compared to conventional security models because the prevalent concept of statehood goes beyond territory, nationality, and population markers in the cyber space. This way, cybersecurity issues will often lack a clear attribution and motivation. This has contributed to cyber warfare becoming significantly offense-dominant coupled with the absence of an effective deterrence [8]. When this is contextualised with rising challenges posed by the interjurisdictional nature of cybersecurity incidents, this means that most states will have a limited capacity to investigate and incarcerate cyber criminals. Hence, it is important for organisations and governments alike to have an accurate appraisal of the risks posed by state and state-sponsored attackers to then adopt an effective offensive orientation as opposed to have a reactive one. Therefore, with the increased adoption of digital services and technologies, state and state-sponsored are capitalising on jurisdictional challenges by national systems and jurisdictional opportunities presented by the internet [8]. Hence, because of the prevalent fragmentation of the law at different levels, this is contributing to the proliferation of state and state-sponsored attackers because of the relatively low possibility of detection and difficulty in attribution which can be used for the administration of justice.

*B. Cyber-kinetic attacks, cyberwarfare, and critical infrastructure*

As outlined by Sigholm, the cyberspace consists of unique attributes that makes it an effective medium for crime, espionage, crime, and military aggression [9]. These include, legal ambiguity, low entry costs, and the asymmetric nature. This is furthered by Jang-Jaccard and Nepal who assert that while societies, economies, and critical infrastructure are significantly reliant on computer networks and information technology innovations, cyber attacks are becoming more attractive and potentially more severe, a trend that is furthered by continued reliance on information technology [10]. Conversely, what this transition has meant for the transition of crime to the cyberspace is effectively captured by Jang-Jaccard and Nepal that unlike traditional crime that required significant levels of investment and effort, a cyber criminal will only require few expenses, the core being an internet connection and a computer [10]. In addition to this, cyber criminals are not confined by distance and geography making it difficult to prosecute and identify them. In addition to this, the main differences between cyber criminals are their inherent motivation and not the resources they have access to, skill level, and other attributes [11]. This makes critical the assertion by Sigholm that the growing importance of the cyberspace and increased ideological and attitudinal differences of threat actors is contributing to the increased appeal and utilisation of the cyberspace for dispute [9]. As such, this establishes the increased operationalisation of the cyberspace by rogue individuals or state-sanctioned or state-condoned parties to purposefully wreak havoc on another state's critical infrastructure albeit for different purposes key among them being disruption of operations or espionage.

Cyber-kinetic attacks are the attacks that target cyber-physical systems to cause indirect or direct injury, environmental impacts, death, or physical damage to the target. In line with the aforementioned impact of underlying intentions, it emerges that it is possible to recruit persons to carry out certain attacks against perceived enemy or competing state(s). This is referred to as strategic sabotage. To this, state and state-sponsored actors while they are considered active threat actors, they are also identified as having varied objectives from typical threat actors. The core motivators are mainly to gather intelligence or to support or advance national interests [11]. Nevertheless, although the concept of sovereignty grants a state the right to defend itself using cyber or kinetic attacks where a cyber-attack physically destroys facilities and threatens people's welfare, a retaliating state will require to have prove that another state is responsible for the attack. This results in two key factors to consider [12]. The first one is that threat actors take great care to avoid detection throughout the lifecycle of the attack and the second one is that while there are associations that have been made between different cyber criminal groups and countries, no country has explicitly declared sanctioning or condoning cyber-attacks. Therefore, while some actors continue to act for or in the benefit of a state, this introduces two key aspects that increase the overall propensity to cyber-kinetic attacks targeting critical infrastructure, the difficulty of proving association and the increased equipment of such teams due to the state linkages.

*C. Key considerations in securing critical infrastructure*

Regarding the aforementioned increased in the utilisation and adoption of critical infrastructure and the resultant increased resilience, the failure to offer effective and efficient cybersecurity controls for critical infrastructure increases the susceptibility of a country to an attack with foreseeably far-reaching impacts on security, safety, economic security, and public health [12]. It is this that establishes the importance of cultivating a country's capacity to identify threats against its critical infrastructure and to effectively safeguard different aspects through approaches such as risk mitigation. Mariani and others presents the key points of focus in securing critical infrastructure, that is, developing a country's capacity to understand the underlying incentives of all stakeholders, including threat actors [13]. It is from this that a country can come up with effective safeguards, protocols, procedures, and practices. On this, it emerges that there are two key factors that set out the core focus areas in safeguarding critical infrastructure, promoting overall uptime and availability of systems and services offered through critical infrastructure solutions, and, promoting the delivery of functional value to all aspects of the modern society.

In the present-day cyber environment, there is a significant shift in the mandate of cybersecurity professionals. That is, the traditional goal of cyber security professionals mainly involved promoting confidentiality, availability, and integrity of IT assets. However, with the increased sophistication, regimentation, structuring, and financing of present-day actors, there is an emerged responsibility among cyber security experts to continually evaluate the environment for threats and vulnerabilities that challenge the safety of critical infrastructure operators and their operational technologies (OT) [14]. However, cybersecurity practices have consistently lagged in addressing OT vulnerabilities and improving cybersecurity challenges associated with inadequate protection of operational systems, connected devices, and control system. As such, it emerged that effective securing of critical infrastructure in present-day society is characterised by meeting the priorities under both OT and IT infrastructure where the former prioritises safety and reliability in operations, and the later prioritises sensitivity and privacy of information [14]. It is this that sets out the promoting of confidentiality and availability as key target factors. Therefore, it emerges that effective safeguarding of critical infrastructure requires the safeguarding of the associated IT and OT aspects where each requires the adoption of a different focus.

There is also a need to account for the impact of technological advancements. Case in point, Daricili and Celik assert that advancements in technology have increasingly centered critical infrastructure in modern life but they have also made modern communities highly vulnerable [15]. On this, the effectiveness and efficiency of management serves as a key indicator for a state's economic development and social welfare. On this, Mariani and others assert that advancements such as increased computing power and the reducing size and costs of components such as processors, memory sticks, and batteries are contributing to the increased blurring of the line between digital and physical worlds [13]. As such, objects that have traditionally been physical are now embedded with digital sensors and actuators linked to central IT networks, and at times to the wider internet. This contributes to the convergence of IT and OT aspects making critical infrastructure potentially accessible to threat actors over the internet. This presents the need to ensure the alignment between critical infrastructure safeguarding approaches and the prevalent state of technology [16].

Finally, the safeguarding of critical infrastructure requires advanced focus on the development and implementation of effective critical information infrastructure protection (CIIP) priorities, policies, and strategies which is a subset of critical infrastructure protection (CIP). Comparatively, CIIP adopts a more global orientation while CIP adopts a more national orientation. This way, it is importance for organisations to develop strong partnerships especially those centered on improving information sharing and exchange performance and capabilities [14]. This is especially apparent when considered within the context of the findings by Izycki and Vianna that while attacks against critical infrastructure are not as frequent, they require some degree of specialised knowledge on how to target the control systems [12]. As such, the increase in the number of countries with cyber offensive capability, will contribute to an increase in attacks against critical infrastructure resulting in the increased risk of cyber-kinetic attacks on critical infrastructure. Case in point, in their evaluation of the security of industry control system (ICS) which refers to the collection of processes automation technologies, Maglaras and others asserts that inadequate investment in the security of ICS has contributed to the insufficient understanding of components such as intelligence electronic devices (IED), input/output (I/O) devices, remote terminal units (RTU) and programmable logic controllers (PLC) [17]. As such, each of these components present a viable channel of infiltration and attaining different objectives such as service disruption or espionage. As such, this establishes the importance of countries to establish cyber norms through multilateral debates. To this, while CIP focuses on securing critical infrastructure based on a country's contextual realities, the adoption of a CIIP focus promotes security and availability outcomes for critical assets since critical information infrastructures (CII) are comprised of ICT process and information control systems which takes a more comprehensive approach. Hence, safeguarding the overall information infrastructure emerges as a key approach to promote the holistic safeguarding of critical infrastructure.

*D. Common tactics employed by state and state-sponsored actors*

As aforementioned that the intentions of an attacker are what will differ and influence several aspects about a cyber-attack. On this, state and state-sponsored will be typically motivated by the pursuance of military, economic, or political interests of the country. To do this, this group of threat actors will use malicious cyber campaigns. Nevertheless, it is also important to highlight that since most state or state-sponsored attacks will entail an element of disruption or espionage, threat actors will foreseeably rely on different cyber-techniques key among them being deployment of malware, distributed denial of service (DDoS) attacks, and advanced persistent threats tactics. Narrowly, the approaches make it possible for attackers to disrupt a critical infrastructure and to establish a long-standing presence in the target's digital environment.

A DDoS attack is narrowly characterised as malicious attempt to disrupt the normal flow of typical tasks such as traffic flow to a service or serve attained through overwhelming a target or the surrounding infrastructure with internet traffic. In doing so, the attacker makes an online service unavailable to users hence disrupting normal carrying out of tasks. For the case of critical infrastructure, target nations or nation agencies will be faced with have their resources flooded with HTTP traffic and requests hence denying access to legitimate users [18]. This way, while it is imperative to promote the overall rate to which critical infrastructure remains active to support different aspects of life such as delivery of government services, DDoS services are designed to cripple critical infrastructure making it difficult, if at all possible, to serve the public. In line with cyber-kinetic attacks, since state and state-sponsored actors will often be well-resourced and sophisticated, DDoS attacks are utilised as a way to significantly damage cyber infrastructure, for instance, disrupting power grids, and shutting down access to websites and servers. A key example of this is the Russian government-sponsored group referred to as Sandworm which focuses on the creation of botnets aimed at causing harm through the deployment of DDoS [18]. This way, it emerges that for state and state-sponsored attackers whose motive is to disrupt critical infrastructure, carrying out DDoS attacks is among the approaches utilised and appropriate attack vector.

Turning the focus to malware, it achieves the same as DDoS from the perspective of disruption although malware can be programmed to carry out different activities throughout a target's system or network including data exfiltration. On this, it is important to highlight that attackers will need to remotely access a system or network from where they can use a command and control (C&C) server to send out instructions to a malware based on the objectives of an attack. Therefore, with malware being characterised as software utilised to disrupt normal computer operations, gain unauthorised access, and to gather sensitive information, state and state-sponsored actors rely on this approach to spy and/or disrupt a country's digital environment [19]. Case in point, from 2011 to 2018, the Russian Federal Security Services launched intrusion campaigns against the US and other global energy sector organisation. Specifically, these campaigns were characterised by the use of the Havex malware which is remote access trojan (RAT) that infected the systems of energy sector networks including exfiltration of process data. In line with the outlined need to establish a connection with a C&C server, the server interacted with RAT to deploy payloads that computes network resources whiles deploying the Open Platforms Communications (OPC) standard to collect information on the prevalent resources and the connected control system devices in the network [20]. Therefore, the malware made it possible to install more malware, and collect information on aspects such as the list of installed programs, system information, VPN configuration files, and email address books. This way, it emerges that malware is not only a key mode of attack in the cyber space, but it is also preferred among state and state-sponsored actors, especially those whose attack requires a form of remote connectivity.

Finally, in the case of APT, it refers to an attack campaign where an attacker or a group seeks to ensure an unauthorised and long-standing presence within a network to mine highly sensitive data. Unlike typical cyber-attacks, targets are well researched and chosen, and the attack will require more resources. As a result, the deployment of APT attacks requires having a significant financial banking and as a result, some attacks will be government-funded to be used as cyber warfare weapons [21]. Also, it is important to highlight that the core characteristics of this form of attack highlights that the main goal of such is to not only collect information but to also remain undetected throughout the attack lifecycle. Case in point, in line with the ongoing Russia/Ukraine conflict, the aforementioned group Sandworm is among other Russian groups capitalising on APT targeting Ukraine. Other examples include InvisiMole, Turla, Callista, and Gamaredon [22]. In the case of Sandworm, the group launched what is referred to as the CaddyWiper campaign which is characterised by the deployment of a data-wiping malware that resulted in Ukrainian organisations having their partition information and user data on drives destroyed. The malware simultaneously leaks information through social media platform Telegram [23]. From this, it is outlined that the core motivations for the attack are diverse, disruption of activities, cyber espionage, and intellectual property theft. Therefore, because of the alignment of the nature of APT attacks with the objectives of different state and state-sponsored groups, this makes APT an ideal channel of attack due to the high success rate and low detection rate.

*E. State and state-sponsored attacks case study*

  *1) Stuxnet*

Designated as a malicious computer work that was first identified in 2010, Stuxnet targets supervisory control and data acquisition (SCADA) systems believed to have caused damage to Iran's nuclear program. As such, it emerged as the first virus to cause the physical destruction of infected devices and ultimately crippling Iran's nuclear program [24]. That is, since it was designed to affect devices with specific configurations, it had a code for a complex man-in-the-middle attack which faked sensor signals and as a result, the nuclear monitoring system would not send alerts or shutdown due abnormal behavior. As such, this establishes the disruptive impact of attacks by state and state-sponsored actors.

Turning the focus to the orchestration of the attack, it is believed that the worm was created by the US and Israel jointly as part of what is referred to as Operation Olympic Games. While the operation still remains covert and still unacknowledged, it is was a strategic sabotage operation relying on cyber disruption as the preferred mode of attack targeting Iranian nuclear facilities [25]. Operation-wise, the worm infiltrated Windows systems through zero-day vulnerabilities such as remote code execution, vulnerabilities in the printer spooler, and the LNK/PIF vulnerability. The malware then accessed user and kernel levels to have its drivers signed without a user's knowledge and to remain undetected. Successful infiltration is then followed by the infection of the Siemens industrial software application files to disrupt how they communicate with each other while also modifying the code on PLC devices [24]. This way, the worm blocks PLC monitors whilst constantly changing a system's frequency and the rotational speed of motors. As such, this proves that most attacks by state and state-sponsored groups will employ complex approaches and within the context of cyber-kinetic attacks, and cyber warfare in general, they will target perceived enemy or competing state. Therefore, cyber-attacks through state and state-sponsored groups while they cannot be claimed by nation states due to the challenge of sovereignty, they can be sanctioned as covert operations to promote the interests of a nation state.

  *2) SolarWinds hack*

Deployed in 2020, the solar winds attack is suspected to have been orchestrated by a hacker group, Nobelium, backed by the Russian government that resulted in the infiltration of US organisations and government agencies that resulted in a series of data breaches. Adopting a more granular looks highlights that, the cyberattack involved the SolarWinds Orion System which is among SolarWind's products focused on delivering IT performance monitoring to its clients [26]. As such, the Orion platform has privileged access to IT systems from where it is possible to obtain system performance and log data [27]. By weaponising this access protocol, Nobelium gained access to systems, networks, and data for a wide array of SolarWinds clientele where it is estimated to have affected more than 30,000 private and public organisations using the Orion platform to manage their IT resources. Therefore, by gaining unauthorised access, the Orion platform served as a backdoor where the hacker group accessed and impersonated users and accounts [27]. From this, it made it possible to access system files disguising the activity as legitimate SolarWinds' activity to evade detection from different safeguards across organisations such as intrusion detection and prevention systems and antivirus software.

In terms of the impacts of the attack, the SolarWinds hack played the role of both espionage and intellectual property theft. In the case of espionage, with government departments such as the Department of Homeland Security being among the core victims, it was later identified that some of the emails had been exfiltrated from their systems. From the perspective of intellectual property theft, it should be noted that with FireEye being credited for the identification of the vulnerability, this came after the company identified that one of its tools used by the red team was missing where an audit revealed the backdoor in the platform that was used to access its systems [28]. In line with the outlined importance of state and state-sponsored to avoid detection throughout the attack lifecycle, it is estimated that the attackers first gained access to the system in September 2019 before being discovered in December 2020 which indicates the attackers had close to 14 months of unrestricted and undetected access to the systems of the host of organisations that use the Orion platform. As such, this indicates the degree of sophistication that state and state-sponsored actors can ingrain in a typical attack.

III. DISCUSSION, RECOMMENDATIONS, AND CONCLUSION

With the increased move towards digital transformation of societies, there is an increased adoption of digital devices, increased connectivity across devices, and the utilisation of the internet as a platform for interaction. In turn, this is promoting both efficiency and availability of different services for a wider range of persons. However, in the same way that government and organisational systems and networks are availed to the market, this transformation is contributing to the rising susceptibility these organisations, institutions, and agencies to cyber criminality. On this, the cyber space has inadvertently become the novel group for disputes among individuals and organisations alike. Following the outlined value of a cyber criminal's intention as a marker of their focus, it is foreseeable that increased geopolitical tensions will foreseeably increase the proportion and severity of state and state-sponsored actors however, the foreseeable transition to cyber-kinetic attacks as the case with Stuxnet, and intellectual property theft as evidenced by the SolarWinds Hack present a key concern especially due to the affordances of the cyberspace, namely, legal ambiguity, low entry costs, and the asymmetric nature coupled with low detection capacity.

Therefore, with intention being the core differentiator of cyber criminals, it is foreseeable that basing safeguards on the intentions of a possible attacker type is an effective way to secure critical infrastructure to guarantee economic and social development outcomes. First, this is by adopting a holistic perspective of critical infrastructure by focusing on both its IT and OT assets and adopting a globalised perception of safeguarding critical infrastructure though increasing focus on CIIP. This goes to show that the cyber environment today cannot be perceived from the perspective of just the functional attributes of different innovations, instead, there is a need to have a wider understanding of the different interacting aspects. From this, cyber realities such as modes of attack can be adequately responded to since the focus is on the safeguarding of the infrastructure and ensuring operational continuity. However, this is significantly challenged by the rise of state and state-sponsored actors who in addition to having access to advanced equipment, tools, and knowledge; they are also innately motivated to promote the interests of the nation. This shows that the entry and increased prevalence of state and state-sponsored actors will foreseeably make securing critical infrastructure more complex. However, incidents by state and state-sponsored actors are not as frequent because of the required degree of specialisation both skill-wise and targeting. Therefore, as societies continue to adopt digital transformation, it is important they improve CIP outcomes through the incorporation of CIIP priorities to ensure critical infrastructure is secure based on the advancement of its internal capabilities and competencies but also with regards to global socio-political realities.